\documentclass[column,showpacs,preprintnumbers,amsmath,amssymb]{revtex4}
\preprint{}
\usepackage{graphicx}
\usepackage{dcolumn}
\usepackage{bm}
\usepackage{mathrsfs}
\begin{document}

\title{Multi-parameter Quantum Magnetometry with Spin States in coarsened measurement reference}
\author{Dong  Xie}
\email{xiedong@mail.ustc.edu.cn}
\affiliation{College of Science, Guilin University of Aerospace Technology, Guilin, Guangxi, P.R. China.}

\author{Chunling Xu}
\affiliation{College of Science, Guilin University of Aerospace Technology, Guilin, Guangxi, P.R. China.}

\begin{abstract}
We investigate the simultaneous estimation of the intensity and the orientation of a magnetic field by the multi-parameter quantum Fisher information matrix. A general expression is achieved for the simultaneous estimation precision of the intensity and the orientation, which is better than the independent estimation precision for the given number of spin states. Moreover, we consider an imperfect measurement device, coarsened measurement reference. For the case of the measurement reference rotating around the $y-$axis  randomly, the simultaneous estimation always performs better than the independent estimation.
For all other cases, the simultaneous estimation precision will not perform better than the independent estimation when the coarsened degree is larger than a certain value.
\end{abstract}

\pacs{03.65.Yz; 03.65.Ud; 42.50.Pq}
\maketitle

\section{Introduction}
Quantum metrology mainly involves obtaining fundamental sensitivity limits and developing strategies to enhance the precision of parameter estimation with quantum resource\cite{lab1,lab2,lab3,lab4}. There are widespread applications about quantum estimation of a single parameter\cite{lab5,lab6,lab7,lab8,lab9}. Recently, simultaneous quantum-enhanced estimation of multiple parameters is becoming more and more interesting, which is drawing more attention.  It is mainly because of the fact that unlike in the quantum single-parameter estimation case, quantum measurements required to attain multi-parameter bounds do not necessarily commute\cite{lab10,lab11}. Multi-parameter estimation also has many important applications, such as, quantum imaging\cite{lab12,lab13,lab14}, microscopy and astronomy\cite{lab15,lab16}, sensor networks\cite{lab17,lab18}. All these tasks go beyond single-parameter estimation. There are a lot of theoretical works\cite{lab19,lab20,lab21,lab22,lab23,lab24,lab25,lab26,lab27,lab28}, which clearly show that simultaneous estimation can be more precise than estimating the parameters individually.

Utilizing quantum resource to improve the estimation precision of magnetic field has draw many attention\cite{lab29,lab30,lab31,lab32,lab33}. A variety of spin systems \cite{lab34,lab35,lab36,lab37,lab38,lab39,lab40,lab41}are currently being used to implement field sensors.
In general, the maximally entangled pure states are required in order to outperform classical devices.
Recently, Filippo Troiani et al.\cite{lab42}, address the single parameter estimation of a magnetic
field, obtained by performing arbitrary measurements on the equilibrium state of an arbitrary spin.
The advantage is that decoherence no longer represents a limiting factor for the equilibrium state.

In this article, we perform the simultaneous estimation of two parameters: the intensity and the orientation of a magnetic field. We derive a general expression of two-parameter quantum Fisher information matrix, and show that estimating multiple parameters simultaneously can be more precise than estimating the parameters individually.

A complete measurement can be divided into two steps: the first step involves setting up a measurement reference and controlling it, and the second step involves utilizing the corresponding projector to perform the final measurement. We call the imperfect appearing in the first step as \textbf{coarsened measurement reference}. In ref.\cite{lab43}, the role of coarsened measurement reference in a single parameter estimation has been investigated. In this article, we study the role of coarsened measurement reference in simultaneous  multi-parameter estimation. For the case of the measurement reference rotating around the $y-$axis,  the simultaneous estimation always perform better than the independent estimation.
For all other cases, the simultaneous estimation precision will not perform better than the independent estimation when the coarsened degree is larger than a certain value.

The rest of this article is arranged as follows. In section II, we briefly introduce the model of multi-parameter quantum magnetometry and derive a general expression of two-parameter quantum Fisher information matrix. In section III, we detail the role of coarsened measurement reference in simultaneous  multi-parameter estimation. Then, we obtain multi-parameter precision with a given observable in section IV.  A conclusion and outlook are presented in section V.
\section{Multi-parameter Quantum Magnetometry }
We consider that a finite spin system with equispaced energy levels can be described as spin operator $\mathbf{S}$, placed in an external magnetic field\cite{lab42}. The field depends on multiple unknown parameters \textbf{$\vec{\lambda}$}=\{$\lambda_1$,$\lambda_2$,...,$\lambda_i$,...,$\lambda_m$\} both in the intensity and the orientation. The system Hamiltonian is represented as
\begin{eqnarray}
\mathcal{H}=\omega(\sin \theta S_x+\cos \theta S_z)=\omega \mathbf{\hat{n}_Z}\cdot\mathbf{S}=\omega S_Z
\end{eqnarray}
where the direction $\mathbf{\hat{n}_Z}=({\sin \theta\cos\varphi,\sin \theta\cos\varphi, \cos\theta})$ and the energy gap $\omega$ (intensity) are known functions of $\lambda_i$. In order to simplify the equations, and without loss of generality, we consider $\varphi=0$ in the following content.

The density operator of spin system in equilibrium with a heat bath at a temperature T is given by
\begin{eqnarray}
\rho_{\vec{\lambda}}=\sum_{M_Z=-S}^S \frac{e^{-\delta M_Z }}{\mathcal{Z}}|M_Z\rangle\langle M_Z|,
\end{eqnarray}
where $\mathcal{Z}=\sum_{M_Z=-S}^Se^{-\delta M_Z }$ denotes the partition function, $\delta=\omega/k_BT$($\hbar=1$) represents the ratio between the Hamiltonian and the thermal energy scales, and $|M_Z \rangle=e^{-iS_y\theta}|M_z\rangle$ are the
eigenstates of $S_Z$.

The quantum Fisher information matrix $F(\vec{\lambda})$\cite{lab44,lab45,lab46,lab47} has
matrix elements,
\begin{eqnarray}
F_{\lambda_i\lambda_j}(\vec{\lambda})=\frac{1}{2}\textmd{tr}[\rho_{\vec{\lambda}}(L_{\lambda_i}L_{ \lambda_j}+L_{\lambda_j}L_{\lambda_i })],
\end{eqnarray}
where the symmetric logarithmic  derivative (SLD) $L_{\lambda_i}$ satisfies the equation $\frac{1}{2}(\rho_{\vec{\lambda}} L_{\lambda_i}+L_{\lambda_i}\rho_{\vec{\lambda}})=\partial\rho_{\vec{\lambda}}/\partial L_{\lambda_i}$. Using  projection of the density operator derivative on the Hamiltonian eigenstates, the expression of the SLD are given by

\begin{eqnarray}
L_{\lambda_i}=2\sum_{M_Z,M_Z'}\frac{\langle M_Z|\partial_{\lambda_i}|M_{Z}'\rangle}{\frac{e^{-\delta M_Z }}{\mathcal{Z}}+\frac{e^{-\delta M_{Z}' }}{\mathcal{{Z'}}}}|M_Z\rangle\langle M_{Z}'|\\
=2\frac{\partial\theta}{\partial \lambda_i}\tanh(\delta/2)S_X+\frac{\partial\delta}{\partial \lambda_i}(\langle S_Z\rangle- S_Z).
\end{eqnarray}
Obviously, $L_{\lambda_i}$ corresponding to the different parameters don't commute. This does not immediately imply that it is impossible to simultaneously extract information on all parameters with precision matching that of the separate scenario for each. We find that $\textmd{tr}[\rho_{\vec{\lambda}}[L_{\lambda_i},L_{\lambda_j}]]=0$, which can saturate the multi-parameter quantum information Cram\'{e}r-Rao(CR) bound\cite{lab48,lab49},
\begin{equation}
 \textmd{Cov}(\vec{\lambda})\equiv F^{-1},
\end{equation}
where $\textmd{Cov}(\vec{\lambda})$ refers to the covariance matrix for a locally
unbiased estimator $\vec{\lambda}$, $\textmd{Cov}(\vec{\lambda})_{jk}=\langle(\widetilde{\lambda}_j-\lambda_j)(\widetilde{\lambda}_k-\lambda_k)\rangle$.

Simple communication can show that $\textmd{Cov}(\vec{\lambda})\equiv \infty$ for $m>2$ parameters. This shows that it can not obtain any information of $m>2$ parameters simultaneously.
Considering balanced cost, we can achieve the simultaneous estimation of two parameters
\begin{eqnarray}
\Delta^2\lambda_1+\Delta^2\lambda_2=\textmd{tr}[F^{-1}]=\frac{[({\frac{\partial \delta}{\partial \lambda_1}})^2+({\frac{\partial \delta}{\partial \lambda_2}})^2](\langle S_Z^2\rangle-\langle S_Z\rangle^2)+[({\frac{\partial \theta}{\partial \lambda_1}})^2+({\frac{\partial \theta}{\partial \lambda_2}})^2]4\tanh^2(\delta/2)\langle S_X^2\rangle}{4\tanh^2(\delta/2)\langle S_X^2\rangle(\langle S_Z^2\rangle-\langle S_Z\rangle^2)(\frac{\partial\theta}{\partial \lambda_2}\frac{\partial\delta}{\partial \lambda_1}-\frac{\partial\theta}{\partial \lambda_1}\frac{\partial\delta}{\partial \lambda_2})^2}\\
=\frac{({\frac{\partial \delta}{\partial \lambda_1}})^2+({\frac{\partial \delta}{\partial \lambda_2}})^2}{4\tanh^2(\delta/2)\langle S_X^2\rangle(\frac{\partial\theta}{\partial \lambda_2}\frac{\partial\delta}{\partial \lambda_1}-\frac{\partial\theta}{\partial \lambda_1}\frac{\partial\delta}{\partial \lambda_2})^2}+\frac{({\frac{\partial \theta}{\partial \lambda_1}})^2+({\frac{\partial \theta}{\partial \lambda_2}})^2}{(\langle S_Z^2\rangle-\langle S_Z\rangle^2)(\frac{\partial\theta}{\partial \lambda_2}\frac{\partial\delta}{\partial \lambda_1}-\frac{\partial\theta}{\partial \lambda_1}\frac{\partial\delta}{\partial \lambda_2})^2},
\end{eqnarray}
Where $\langle S_Z\rangle=\frac{1}{2}\coth(\frac{\delta}{2})-(S+1/2)\coth[(1/2+S)\delta]$, $\langle S_Z^2\rangle=S(S+1)+\coth(\frac{\delta}{2})\langle S_Z\rangle$ and $\langle S_X^2\rangle=\frac{1}{2}[S(S+1)-\langle S_Z^2\rangle]$.
The first term in Eq.(8), referred to as quantum, depends on the changes of the field direction $\mathbf{\hat{n}_Z}$ and is inversely proportional to the fluctuations of the transverse spin components. The second term, referred to as classical, is inversely proportional to the fluctuations in the longitudinal spin projection, which comes from the incoherent mixture of the Hamiltonian eigenstates $|M_Z\rangle$.

And we find that when $(\frac{\partial\theta}{\partial \lambda_2}\frac{\partial\delta}{\partial \lambda_1}-\frac{\partial\theta}{\partial \lambda_1}\frac{\partial\delta}{\partial \lambda_2})=0$, the uncertainty of simultaneous estimation is infinity. Namely, no any information can be obtained. It is due to that $\theta$ and $\delta$ have similar form. For example,
$\theta=\lambda_1+\lambda_2$ and $\delta=\frac{\lambda_1+\lambda_2}{k_B T}$. The value of $\lambda_1$ and $\lambda_2$ can not be obtained from $\theta$ and $\delta$.
\section{coarsened measurement reference in simultaneous  multi-parameter estimation}
Coarsened measurement includes not only the coarsened measurement precision but also the coarsened reference\cite{lab50}
The coarsened reference can exert a more significant influence in a single parameter quantum metrology than the coarsened measurement precision\cite{lab43}. We investigate the simultaneous estimation of intensity $\omega$ and direction $\theta$ in coarsened measurement reference.

Firstly,  we consider that the measurement reference basis can randomly rotate around the $z-$axis with coarsened degree $\eta$.
The influence of coarsened measurement reference can be expressed in the density matrix,
\begin{eqnarray}
\rho_{\theta, \omega}=\sum_{M_Z=-S}^S \frac{e^{-\delta M_Z }}{\mathcal{Z}}\int_{-\infty}^\infty \chi_\eta(\phi)e^{-iS_z\phi}|M_Z\rangle\langle M_Z|e^{iS_z\phi},
\end{eqnarray}
where $\chi_\eta(\phi)$ denotes the normalized Gaussian kernel
\begin{eqnarray}
\chi_\eta(\phi)=\frac{1}{\sqrt{2\pi}\eta}\exp(-\frac{\phi^2}{2\eta^2}).
\end{eqnarray}

In order to obtain a simple analytical expression, without loss of generality, we only consider the spin system with two levels (S=1/2). The eigenvectors  of $S_z$ are described by $(|0\rangle,|1\rangle)$.
By a calculation, we can obtain the detail expression of  density matrix in the coarsened measurement reference

 \[
 \rho_{\theta, \omega}= \left(
\begin{array}{ll}
\ p_1\cos^2\frac{\theta}{2}+p_2\sin^2\frac{\theta}{2},\ (p_1-p_2)e^{-\frac{\eta^2}{2}}\frac{\sin\theta}{2} \\
(p_1-p_2)e^{-\frac{\eta^2}{2}}\frac{\sin\theta}{2},\ p_2\cos^2\frac{\theta}{2}+p_1\sin^2\frac{\theta}{2}
  \end{array}
\right ),
\]
\begin{equation}
\end{equation}
where $p_1=\frac{e^{-\delta}}{\mathcal{Z}}$ and $p_2=\frac{e^{\delta }}{\mathcal{Z}}$.
For a two-dimensional system, the multi-parameter QFI matrix is expressed by\cite{lab51}
\begin{equation}
 F_{Q_{ij}}=(\partial_{x_i}{\textbf{r}})\cdot(\partial_{x_j}{\textbf{r}})+\frac{(\textbf{r}\cdot\partial_{x_i}\textbf{r})(\textbf{r}\cdot{\partial_{x_j}\textbf{r})}}{1-|\textbf{r}|^2},
\end{equation}
where $\textbf{r}$ denotes the Bloch vector of a density matrix.
The Bloch vector of $\rho_{\theta, \omega}$ is described by $\mathbf{r}=\left((p_1-p_2)\gamma\sin\theta, 0, (p_2-p_1)\cos\theta\right)$ with $\gamma=e^{-\frac{\eta^2}{2}}$, substituted into Eq.(12) to obtain the multi-parameter QFI matrix
 \[
F({\theta, \omega})= \left(
\begin{array}{ll}
\ \frac{1}{4}\tanh^2(\frac{\delta}{2})[4\gamma^2\cos^2\theta+4\sin^2\theta -\frac{(\gamma^2-1)^2\tanh^2(\frac{\delta}{2})\sin^2(2\theta)}{-1+\tanh^2(\frac{\delta}{2})\cos^2\theta+\tanh^2(\frac{\delta}{2})\gamma^2\sin^2\theta}],\
\frac{\alpha(\gamma^2-1)\tanh(\frac{\delta}{2})\sin(2\theta)}{-1+\tanh^2(\frac{\delta}{2})(\cos^2\theta+\gamma^2\sin^2\theta)} \\
\frac{\alpha(\gamma^2-1)\tanh(\frac{\delta}{2})\sin(2\theta)}{-1+\tanh^2(\frac{\delta}{2})(\cos^2\theta+\gamma^2\sin^2\theta)} ,\ \ \ \ \ \ \ \ \ \ \ \ \ \ \ \ \ \ \ \ \ \ \ \ \ \ \ \ \ \ \ \ \ \ \ \ \ \ \ \ \ \ \ \ \ \ \ \ \ \   \ \  \frac{4\alpha^2}{\tanh^2(\frac{\delta}{2})-\frac{1}{\cos^2\theta+\gamma^2\sin^2\theta}}
  \end{array}
\right ),
\]
\begin{equation}
\end{equation}
where $\alpha=\frac{2e^{2\delta}}{k_BT(1+e^{2\delta})^2}$.

From the above equation, the simultaneous estimation precision of $\theta$ and $\omega$ is achieved
\begin{equation}
\Delta^2\theta_s+\Delta^2\omega_s=\textmd{tr}[F_{\theta, \omega}^{-1}]=\frac{4\alpha^2+\tanh^2(\frac{\delta}{2})+[4\alpha^2-(2\tanh^2(\frac{\delta}{2})-1)\tanh^2(\frac{\delta}{2})]\gamma^2+[\tanh^2(\frac{\delta}{2})-4\alpha^2](\gamma^2-1)\cos(2\theta)}{8\gamma^2\alpha^2\tanh^2(\frac{\delta}{2})}.
\end{equation}

The independent estimation precision of $\theta$ and $\omega$ are given by
\begin{eqnarray}
&\Delta^2\theta|_i+\Delta^2\omega|_{i}=2(F_{11}^{-1}+F_{22}^{-1})\\
&=2(\frac{1}{\frac{1}{4}\tanh^2(\frac{\delta}{2})(4\gamma^2\cos^2\theta+4\sin^2\theta -\frac{(\gamma^2-1)^2\tanh^2(\frac{\delta}{2})\sin^2(2\theta)}{-1+\tanh^2(\frac{\delta}{2})\cos^2\theta+\gamma^2\tanh^2(\frac{\delta}{2})\sin^2\theta})}+\frac{1}{\frac{4\alpha^2}{\tanh^2(\frac{\delta}{2})-\frac{1}{\cos^2\theta+\gamma^2\sin^2\theta}}})
\end{eqnarray}

In the perfect measurement reference, $\gamma=1$, we can find that $\Delta^2\theta_s+\Delta^2\omega_s$=$1/2(\Delta^2\theta|_i+\Delta^2\omega|_{i})$. Namely, given the fixed number of spin states, the simultaneous estimation can perform better than the independent estimation. However, in coarsened measurement reference, in particular, when $\gamma=0$, $\Delta^2\theta_s+\Delta^2\omega_s=\infty\gg\Delta^2\theta|_i+\Delta^2\omega|_{i}$.
In another word, the simultaneous estimation precision will not perform better than the independent estimation when the coarsened degree is larger than a certain value. This can be shown obviously in Fig. 1.
\begin{figure}[h]
\includegraphics[scale=1]{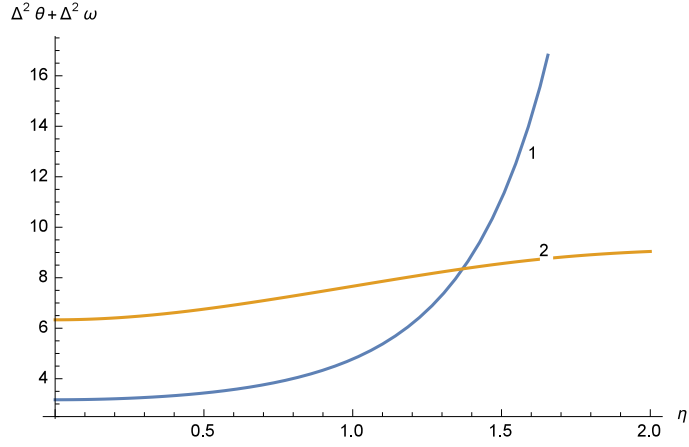}
 \caption{\label{fig.1} The line 1 represents that the simultaneous estimation uncertainty of $\omega$ and $\theta$ changes with coarsened degree $\eta$. The line 2 represents the case of independent estimation. The parameters are given: $\alpha=1$, $\theta=\pi/3$ and $\tanh^2\frac{\delta}{2}=1/3$. }
 \end{figure}

Secondly,  we consider that the measurement reference basis can randomly rotate around the $x-$axis with coarsened degree $\eta$.
The influence of coarsened measurement reference can be expressed in the density matrix,
\begin{eqnarray}
\rho_{\theta, \omega}=\sum_{M_Z=-S}^S \frac{e^{-\delta M_Z }}{\mathcal{Z}}\int_{-\infty}^\infty \chi_\eta(\phi)e^{-iS_x\phi}|M_Z\rangle\langle M_Z|e^{iS_x\phi}.
\end{eqnarray}
For two dimensional system, the corresponding expression of the density matrix is described by
 \begin{eqnarray}
\rho_{\theta, \omega}=\frac{1+\mathbf{r}\cdot\overrightarrow{\mathbf{\sigma}}}{2}
\end{eqnarray}
where the Bloch vector $\textbf{r}=(p_1-p_2)(\sin\theta,0 ,\gamma\cos\theta)$, the Pauli vector$\overrightarrow{\mathbf{\sigma}}=(\sigma_x,\sigma_y,\sigma_z)$.
Then, using  Eq.(12), we can achieved the multi-parameter QFI matrix like the Eq.(13).

 \[
F({\theta, \omega})= \left(
\begin{array}{ll}
\ \frac{1}{4}\tanh^2(\frac{\delta}{2})(4\cos^2\theta+4\gamma^2\sin^2\theta -\frac{\tanh^2(\frac{\delta}{2})(\gamma^2-1)^2\sin^2(2\theta)}{-1+\tanh^2(\frac{\delta}{2})\gamma^2\cos^2\theta+\tanh^2\frac{\delta}{2}\sin^2\theta}),\
\frac{\alpha\tanh(\frac{\delta}{2})(\gamma^2-1)\sin(2\theta)}{-1+\tanh^2(\frac{\delta}{2})(\gamma^2\cos^2\theta+\sin^2\theta)} \\
\frac{\alpha\tanh(\frac{\delta}{2})(\gamma^2-1)\sin(2\theta)}{-1+\tanh^2(\frac{\delta}{2})(\gamma^2\cos^2\theta+\sin^2\theta)} ,\ \ \ \ \ \ \ \ \ \ \ \ \ \ \ \ \ \ \ \ \ \ \ \ \ \ \ \ \ \ \ \ \ \ \ \ \ \ \ \ \ \ \ \ \ \ \ \ \ \   \ \  \frac{4\alpha^2}{\tanh^2(\frac{\delta}{2})-\frac{1}{\gamma^2\cos^2\theta+\sin^2\theta}}
  \end{array}
\right ).
\]
\begin{equation}
\end{equation}
As the same discussion in the above case of rotating around $z-$axis,  the simultaneous estimation precision will not perform better than the independent estimation when the coarsened degree is larger than a certain value.

Thirdly, the measurement reference basis can randomly rotate around the $y-$axis with coarsened degree $\eta$.
For two dimensional system, the corresponding expression of the density matrix is described by
 \begin{eqnarray}
\rho_{\theta, \omega}=\frac{1+\mathbf{r}\cdot\overrightarrow{\mathbf{\sigma}}}{2}
\end{eqnarray}
where the Bloch vector $\textbf{r}=(p_1-p_2)\gamma(\cos\theta,0 ,-\sin\theta)$.

In this situation, the multi-parameter QFI matrix can be simply expressed
 \[
F({\theta, \omega})= \left(
\begin{array}{ll}
\gamma^2\tanh^2(\frac{\delta}{2}),\ \ \ 0 \\
0, \ \ \ 4\alpha^2\gamma^2(1+\frac{\gamma^2\tanh^2(\frac{\delta}{2})}{1-\gamma^2\tanh^2(\frac{\delta}{2})})
  \end{array}
\right ).
\]
\begin{equation}
\end{equation}

As a result, we achieve that  $\Delta^2\theta_s+\Delta^2\omega_s$=$1/2(\Delta^2\theta|_i+\Delta^2\omega|_{i})$ for different value of coarsened degree. That is to say, for the case of the measurement reference rotating around the $y-$axis randomly, the simultaneous estimation always perform better than the independent estimation.
\section{multi-parameter precision with a given observable}
In the above section, we use the QFI matrix to theoretically achieve the optimal bound in coarsened measurement reference. It may not be very appealing from an experimental perspective. Hence, let us discuss some realistic measurements to support the results in the above section.

In particular, we consider a  set of POVMs in two dimensional spin system: $\Pi_1=\frac{1}{2}|0\rangle\langle0|$, $\Pi_2=\frac{1}{4}(|0\rangle+|1\rangle)(\langle0|+\langle1|)$, and $\Pi_3=1-\Pi_1-\Pi_2$.
Due to that the results of rotating around the axis $z$ and axis $x$ are similar, we only discuss about two kinds of coarsened measurement reference: rotating around the $z-$  and $y-$ axes .

Firstly, when the measurement reference basis can randomly rotate around the $z-$axis with coarsened degree $\eta$,
the measurement probability can be described by
\begin{eqnarray}
P_1=\textmd{tr}[\Pi_1\sum_{M_Z=-S}^S \frac{e^{-\delta M_Z }}{\mathcal{Z}}\int_{-\infty}^\infty \chi_\eta(\phi)e^{-iS_z\phi}|M_Z\rangle\langle M_Z|e^{iS_z\phi}]\nonumber\\
=\frac{1}{2}(p_1\cos^2\frac{\theta}{2}+p_2\sin^2\frac{\theta}{2}),
\end{eqnarray}
\begin{eqnarray}
P_2=\textmd{tr}[\Pi_2\sum_{M_Z=-S}^S \frac{e^{-\delta M_Z }}{\mathcal{Z}}\int_{-\infty}^\infty \chi_\eta(\phi)e^{-iS_z\phi}|M_Z\rangle\langle M_Z|e^{iS_z\phi}]\nonumber\\
=(p_1-p_2)\frac{\gamma}{4}\sin\theta+\frac{1}{4},
\end{eqnarray}
\begin{eqnarray}
P_3=1-P_1-P_2,
\end{eqnarray}
where $p_1=\frac{e^{-\delta}}{\mathcal{Z}}$ and $p_2=\frac{e^{\delta }}{\mathcal{Z}}$.
We can obtain the analytical estimation precision by substituting above probability equation  into the following classical Fisher information matrix

 \[
F({\theta, \omega})_c= \left(
\begin{array}{ll}
\sum_{i=1}^3\frac{(\partial P_i/\partial \theta)^2}{P_i},\ \ \ \sum_{i=1}^3\frac{(\partial P_i/\partial \theta)(\partial P_i/\partial \omega)}{P_i} \\
 \sum_{i=1}^3\frac{(\partial P_i/\partial \omega)(\partial P_i/\partial \theta)}{P_i}, \ \ \ \sum_{i=1}^3\frac{(\partial P_i/\partial \omega)^2}{P_i})
  \end{array}
\right ).
\]

However, the result is too lengthy. We numerically reveal the final results, as shown in Fig. 2.
We can see that the coarsened measurement reference makes the simultaneous estimation lose the advantage over independent estimation.
\begin{figure}[h]
\includegraphics[scale=1]{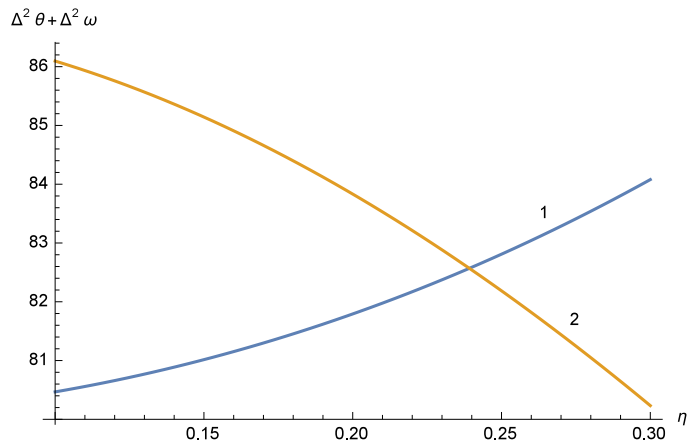}
 \caption{\label{fig.2} The line 1 represents that the simultaneous estimation uncertainty of $\omega$ and $\theta$, obtained by the POVMs, changes with coarsened degree $\eta$ from rotating around the $z-$axis. The line 2 represents the case of independent estimation. The parameters are given: $\alpha=1$, $\theta=\pi/3$ and $p_1=1/3$. }
 \end{figure}
 In a similar way, we discuss the case of rotating around the $y-$axis. As shown in Fig. 3, the simultaneous estimation still has the advantage over independent estimation. Therefore, a given practical measurement operator independent of parameters can show the similar results about the role of coarsened reference as the optimal measurement operator in the above section.
\begin{figure}[h]
\includegraphics[scale=1]{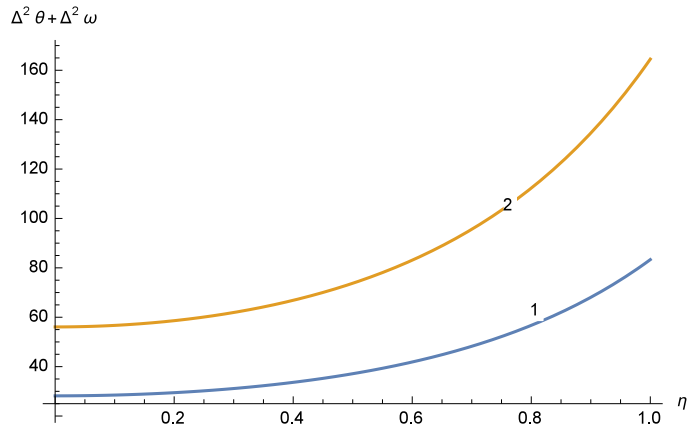}
 \caption{\label{fig.3} The line 1 represents that the simultaneous estimation uncertainty of $\omega$ and $\theta$, obtained by the POVMs, changes with coarsened degree $\eta$ from rotating around the $y-$axis. The line 2 represents the case of independent estimation. The parameters are arranged as: $\alpha=1$, $\theta=\pi/3$ and $\tanh^2\frac{\delta}{2}=1/3$. }
 \end{figure}
\section{conclusion and outlook}
We have investigated the multi-parameter quantum estimation in a magnetic field with a spin at equilibrium. Only two parameters in our model can be simultaneously measured, and the corresponding expression of two parameters estimation precision is achieved. What's more, the role of coarsened measurement reference in multi-parameter Quantum Magnetometry with Spin States have been studied. We utilize the quantum and classic Fisher matrix to obtain the analytical and numerical estimation precisions of two parameters: for the case of the measurement reference rotating around the $y-$axis randomly, the simultaneous estimation always performs better than the independent estimation; for all other cases, the simultaneous estimation precision will not perform better than the independent estimation when the coarsened degree is larger than a certain value. It means that in general,  the independent is more resistant to the interference of the coarsened reference than the simultaneous estimation. Hence, it is necessary to reduce the uncertainty of the coarsened measurement in the simultaneous estimation.

Our investigation will excite the further study of the role of coarsened measurement precision (an imperfect appearing in the second step of a complete measurement) in multi-parameter quantum magnetometry with spin states at equilibrium.
\section*{Acknowledgement}
 This research was supported by the National
Natural Science Foundation of China under Grant No. 11747008 and Guangxi Natural Science Foundation 2016GXNSFBA380227.

\end{document}